\documentclass[12pt]{spieman}  % 12pt font required by SPIE;
\usepackage{amsmath,amsfonts,amssymb}
\usepackage{graphicx}
\usepackage{setspace}
\usepackage{tocloft}
\usepackage{lineno}
\usepackage[dvipsnames]{xcolor}

\newcommand{\ixpe}{\mbox{\em IXPE\/}}
\def \degr {\hbox{$^\circ$}}

\title{Proof of Concept Measurements of Laterally Graded Multilayers for Soft X-ray Spectropolarimetry}

\author[a,*]{Sarah Heine}
\author[a]{Herman L.\ Marshall}
\author[a]{Alan Garner}
\author[b]{Eric Gullikson}
\author[a]{Nithya Kothnur}

\affil[a]{Kavli Institute for Astrophysics and Space Research, Massachusetts Institute of Technology, Cambridge, MA, USA}
\affil[b]{Lawrence Berkeley National Laboratory, Berkeley, CA, USA}

\cftpagenumbersoff{figure}
\cftpagenumbersoff{table} 
\begin{document} 
\maketitle

\begin{abstract}

\end{abstract}

% Include a list of up to six keywords after the abstract

% Include email contact information for corresponding author
{\noindent \footnotesize\textbf{*}Sarah Heine, \linkable{saraht@mit.edu}}

% \begin{spacing}{2}   % use double spacing for rest of manuscript

\section{Introduction}
\label{sect:intro}  % \label{} allows reference to this section
Polarimetry provides two extra dimensions that have been underutilized in X-ray astronomy due to lack of sensitive instrumentation.  Generally, light can be polarized when it is emitted in or passes through areas with strong magnetic fields or scatters, e.g.\ off electrons in accretion disk atmospheres.  There are many examples of the value of optical and radio polarimetry such as finding hidden broad line regions in quasars \cite{1985ApJ...297..621A}, demonstrating that jets and supernova remnants accelerate electrons to very high energies \cite{1959AJ.....64Q.339M,2019ARA&A..57..467B}, and examining the emission regions of pulsar magnetospheres \cite{radha}.  If included, polarization can elucidate the underlying physics and geometry of the emission region much better than non-polarimetric observations.

The Imaging X-ray Polarimetry Explorer (\ixpe \cite{ixpe}) has demonstrated that 
X-ray polarimetry is now a valuable probe of physics in the most extreme environments: the magnetic fields of pulsars \cite{2022NatAs...6.1433D,2022ApJ...940...70M} and magnetars \cite{ixpe_4u0142,ixpe_1708}, the relativistic jets of active galactic nuclei \cite{Liodakis2022,DiGesu2022} and gamma-ray bursts \cite{ixpe_grb211009a}, and the accretion disks around black holes in galactic nuclei \cite{2022MNRAS.516.5907M,2023MNRAS.519.6138M} and in X-ray binaries \cite{ixpe_cygx1}. Despite the great success of IXPE, its sensitivity is limited to 2--8 keV, completely excluding  the soft X-ray band below 1 keV. The additional spectral features typical of the soft X-ray band cannot yet be examined polarimetrically, giving us an incomplete view of these extreme environments. 

\begin{figure}[H]
\begin{center}
\begin{tabular}{c} %% tabular useful for creating an array of images 
\includegraphics[height=5.5cm]{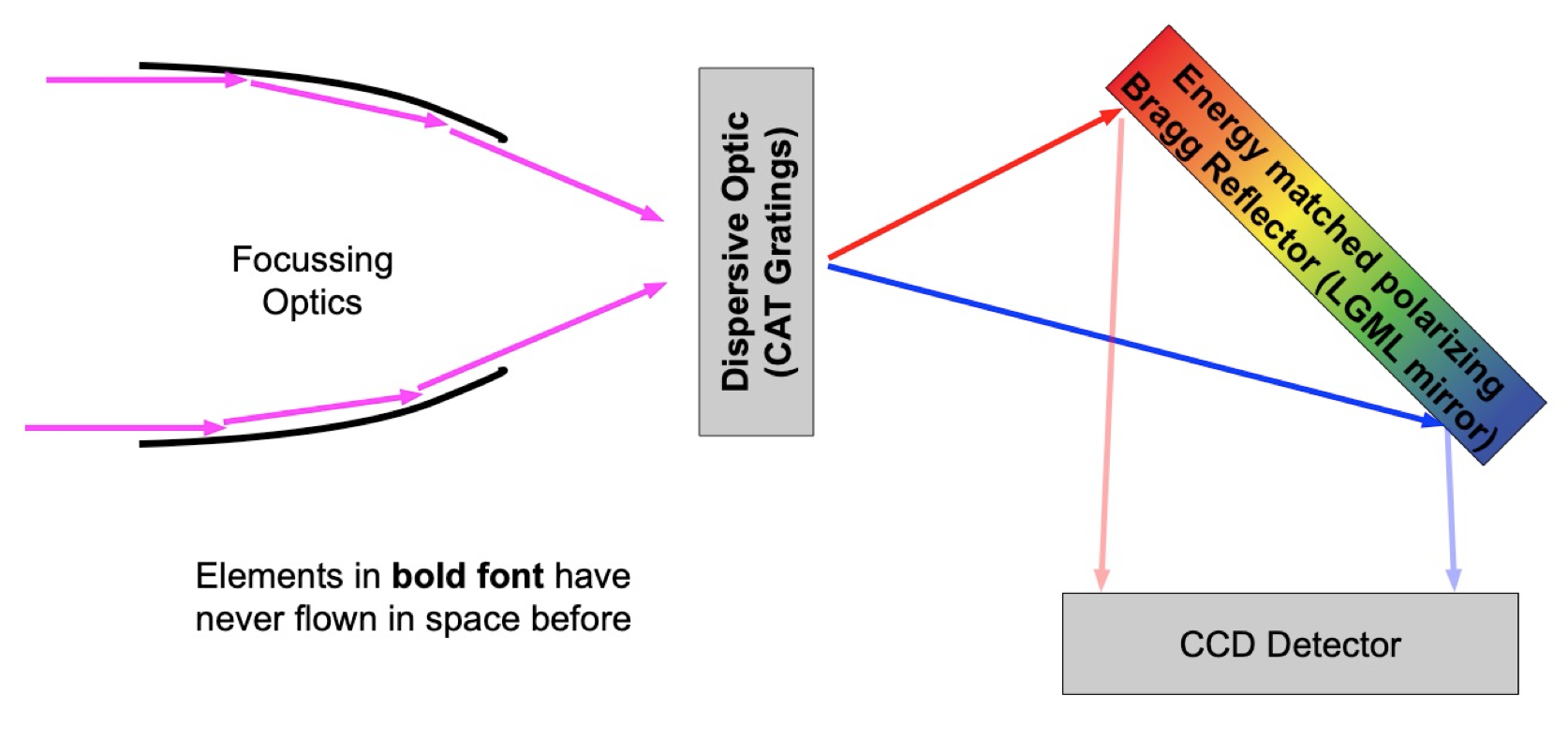}
\end{tabular}
\end{center}
\vspace{-0.2in}
\caption[example] 
%>>>> use \label inside caption to get Fig. number with \ref{}
   { \small 
A block diagram of the optical elements of a single channel (one of three) of the REDSoX polarimeter.}
\label{fig:measurementtechnique}
   \end{figure}

%To do SNTH, put in some REDSoX references
The Rocket Experiment Demonstration of a Soft X-ray Polarimeter (REDSoX\footnote{Used with permission of the Red Sox Baseball Club and Major League Baseball.}) is a NASA-funded sounding rocket mission that will measure polarization of X-ray light in the 200-400 eV band that is being developed by the MIT X-ray polarimetry group.  It is currently in the assembly and test phase \cite{redsox2025,REDSoXSPIE2026}.  REDSoX provides spectropolarimetry measurements by combining precisely aligned optical elements \cite{redsoxjatis}.  Incoming light is incident on the 5 shells of replicated Ni-shell Wolter-I X-ray mirrors, which focus the light with a focal length of 2.5 meters.  The critical angle transmission (CAT) gratings are arranged within the converging beam and disperse the light.  Laterally graded multilayer (LGML) mirrors have a layer spacing that is varied along the surface of the mirror such that the energy of the Bragg peak across the surface of the mirror varies linearly.  The LGMLs are precisely aligned with the CAT gratings such that the energy of the light dispersed by the CAT gratings matches the energy of the Bragg peak at each point along the surface of the LGML.  The 45$\degr$ Bragg reflection off of the LGML reflects light polarized along the surface of the multilayer, which is then measured by a CCD.  There are three channels, with LGMLs oriented at 120\degr\ to each other, which will allow measurement of both the strength and the direction of polarization of the incoming light. A block diagram of one channel is shown in Figure \ref{fig:measurementtechnique}. 

The bandpass of REDSoX is limited by the range of Bragg reflectivity on the multilayer. Our group is working to expand the energy range over which the measurement technique utilized by REDSoX can be employed.  This could bridge the energy gap between the measurements possible utilizing REDSoX (and the newly selected Pioneer, GOSoX \cite{GOSoXSPIE2026}, which will be sensitive in the same waveband as REDSoX), and those made by IXPE (operating 2-8 keV). This can be achieved either by utilizing novel material combinations for multilayer mirrors or by orienting the multilayer with a shallower reflection angle than 45$^\circ$ (the angle employed on the REDSoX payload).

For this work, we will construct a proof of concept test of our polarimeter design and apply it to two multilayers.  We will demonstrate modulation of the reflection strength of polarized light off of a test multilayer under rotation of the polarization angle of the incoming light with respect to the test multilayer.  This test will be repeated for several energies per multilayer to demonstrate that the polarimetry concept works over a broad waveband.  We will test two different multilayers.  A Chromium-Scandium multilayer at 45$^\circ$ incidence in the 200-400 eV energy range like that used by REDSoX, and a WB$_4$C multilayer, designed for use at 30$^\circ$ incidence in the 500-1000 eV range but also useful at 45$^\circ$ incidence.

\section{Experimental setup}
\label{sect:exp_setup}
\begin{figure}[H]
\begin{center}
\begin{tabular}{c} %% tabular useful for creating an array of images 
\includegraphics[width=15cm]{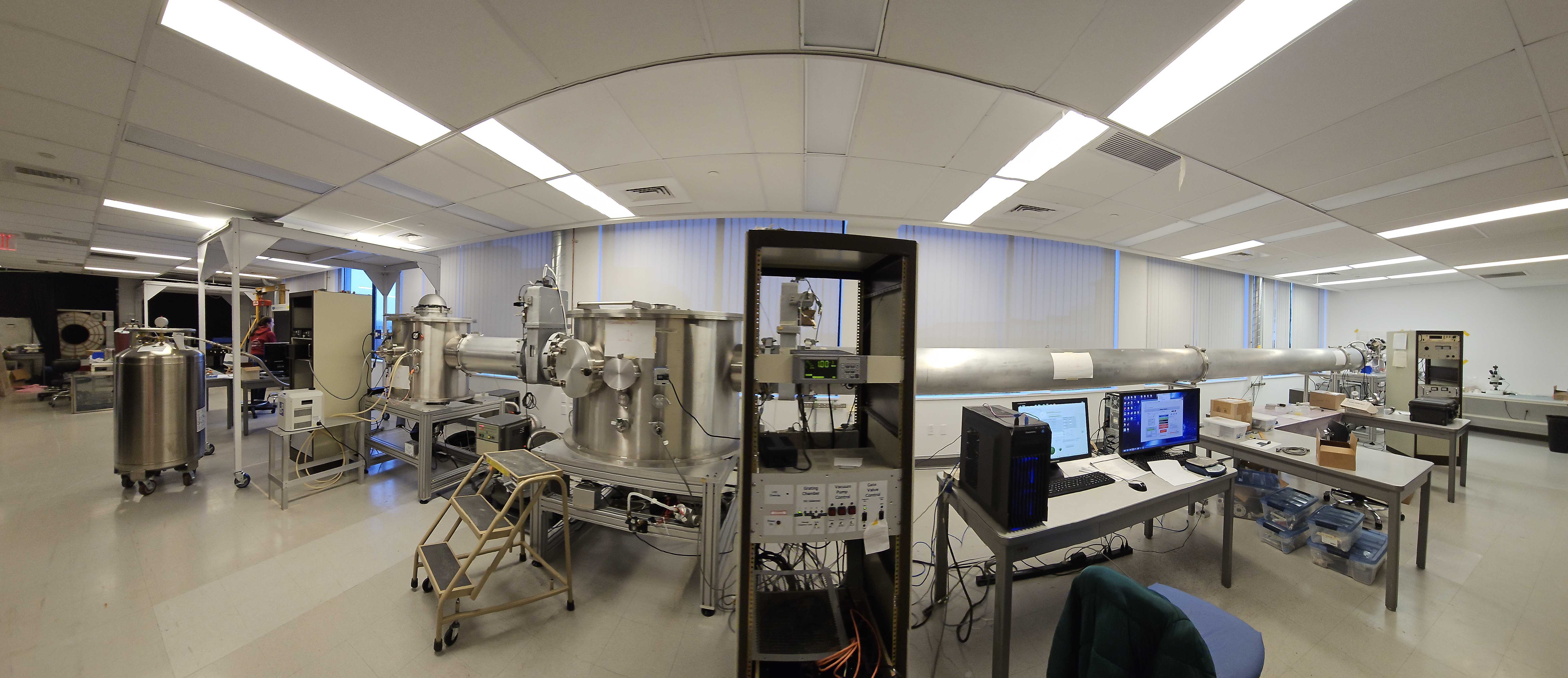}

\end{tabular}
\end{center}
\vspace{-0.2in}
\caption[example] 
%>>>> use \label inside caption to get Fig. number with \ref{}
   { \small \label{fig:beamline}
A photograph of the MIT X-ray Polarimetry Beamline. The source end is on the right side of the picture. The large chamber just left of the rack in the center of the image is the 1.3m-diameter grating chamber, while the chamber on the far left is the 1m-diameter detector chamber. Each chamber can be isolated from the rest of the beamline with pneumatic gate valves.  An antifreeze-based chiller, on a short table, pulls heat from the facility detector on an X-Y stage inside the detector chamber.
}
   \end{figure}

Testing was performed in the MIT polarimetry beamline.  The MIT X-ray Polarimetry Beamline \cite{heine17} is an 11-meter-long X-ray vacuum system (extendable to 17 meters) originally designed for testing the Chandra HETG gratings (Figure \ref{fig:beamline}). It consists of two large cylindrical vacuum chambers: a central chamber, primarily used for mounting gratings or other optics under test, and a second chamber at the far end of the beamline, which houses an in-vacuum detector and can be used for additional detectors or multilayers under test. The second chamber also features a rear flange with an electrically isolating ceramic interface, enabling the attachment of a flange-mounted detector or a small auxiliary chamber for testing detectors.

\begin{figure}[H]
   \begin{center}
   \begin{tabular}{c} %% tabular useful for creating an array of images 
   \includegraphics[width=12 cm]{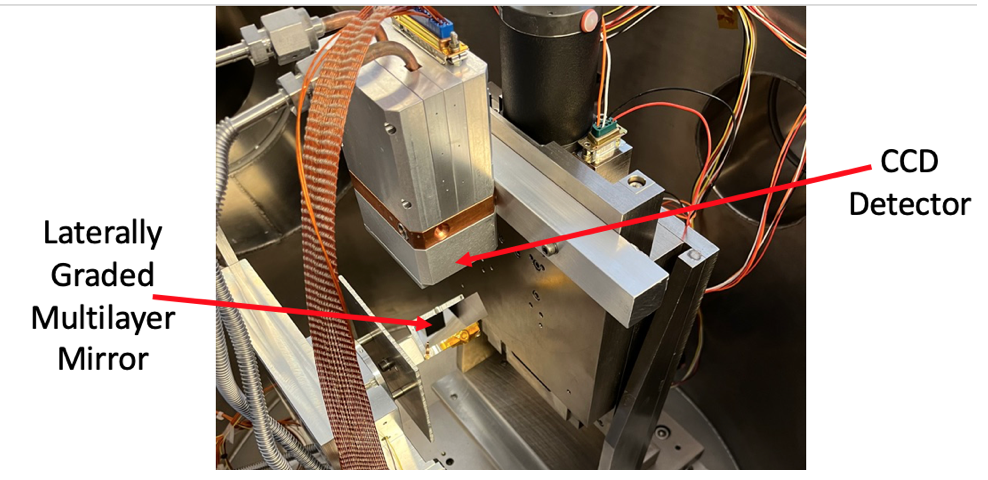}
   \end{tabular}
   \end{center}
   \vspace{-0.2in}
   \caption[example] 
%>>>> use \label inside caption to get Fig. number with \ref{}
   { \small \label{fig:beamlinedet}
The interior of the beamline's detector chamber houses the detector. The Princeton Instruments PI-MTE 1300B CCD detector is mounted on a motorized X-Y stage.  It can be mounted horizontally for use in a direct beam or vertically for testing multilayers as shown here.}
   \end{figure}

The most unique element of the MIT polarimetry beamline is its source.  A Manson source with a carousel of anodes is mounted perpendicular to the beamline.  It produces broad continuum emission with emission lines characteristic of the chosen anode.  The X-rays illuminate a laterally graded multilayer mirror (LGML) on a motorized stage.  The LGML is positioned so that the Bragg peak energy matches the emission line energy being produced, which will reflect only the desired X-ray energy 8 m downstream through a narrow aperture (Figure \ref{fig:beamlinedet}).  This laterally graded multilayer operates at Brewster's angle, allowing only s-polarized X-rays to reflect \cite{2015SPIE.9603E..19M}.  A rotating flange between the source/LGML assembly and the rest of the beamline allows the rotation of the assembly via a motor to control the angle of polarization of the light traveling down the beamline. 

At the entrance of the grating chamber, an aperture stage houses several slits and a 1'' square aperture with a shutter that is synchronized to open during integration by the beamline camera in the detector chamber.  The grating chamber contains stages that allow us to mount gratings or other optical elements, but for this work there were no other optical elements in the chamber.  Given the beamline length and the reflection off the source LGML, this configuration produces a beam that is approximately collimated (though slightly divergent), linearly polarized, and monochromatic (dE/E $\approx$ 0.01).

The beamline camera is a Princeton Instruments MTE 1300B CCD detector, which is installed on a motorized XY stage within the detector chamber.  This detector can be oriented horizontally for tests using a direct beam (most notably grating and detector testing) or vertically to detect light reflecting off of a test multilayer placed in the beam path in the detector chamber.  This test configuration is shown in the schematic in Figure \ref{fig:beamlinedet}.  The detector is 1300 by 1340 imaging array with 20 by 20 micron pixels.  An optical blocking filter is attached to the front of the camera and cooling lines run 10 C antifreeze to the hot side of the integrated thermoelectric cooler from a ThermoCube cooler in a lab rack outside the beamline.  

%This calibration detector can be precisely positioned in and out of the X-ray path, serving as a calibration tool to verify uniform X-ray flux alignment and ensure the expected energy spectrum. It can measure X-ray energy with a full width at half maximum (FWHM) of approximately 80 eV. Once calibration is complete, the MTE detector is retracted to allow measurements of the detector under test. (Figure {\ref{fig:detectorchamber}) 

   % Note: If compiling with LaTeX+dvipdf, please ensure images generated from 
% other software packages have their bounding boxes set correctly.
\begin{figure}[H]
\begin{center}
\begin{tabular}{c} %% tabular useful for creating an array of images 
\includegraphics[width=15 cm]{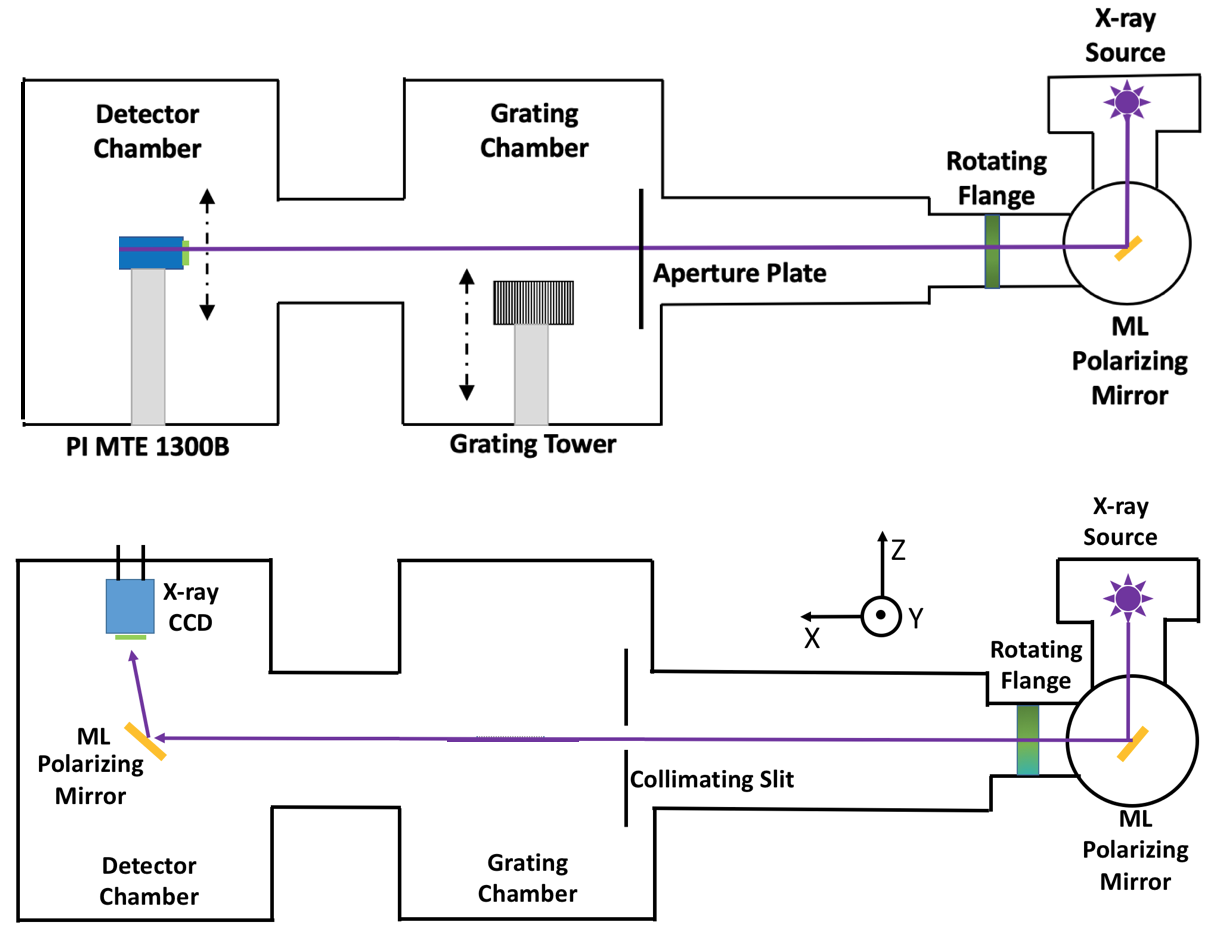}

\end{tabular}
\end{center}
\vspace{-0.2in}
\caption[example] 
%>>>> use \label inside caption to get Fig. number with \ref{}
   { \small \label{fig:beamlineschematic}
Block diagrams of the MIT polarimetry beamline. The beamline is compressed in the diagram for clarity.  The PI MTE 1300B CCD detector is mounted on a stage in the detector chamber.  It can be mounted horizontally for direct beam measurements (top) or vertically for measurements with a test multilayer (bottom).  The source multilayer is on the right side of the schematic closer to the source, while the multilayer under test is on the left side inside the detector chamber.}
   \end{figure} 
\vspace{-0.2in}

\section{Method}
For each test of the LGMLs we selected an anode with an emission line in the region covered by the multilayer under test, and adjusted the source LGML to the position where the Bragg reflectivity peak energy matched that of the emission line.  We began with the detector in the horizontal position shown in the upper diagram in Figure \ref{fig:beamlineschematic}.  We tuned the exact location of the source multilayer by measuring the direct beam and making small adjustments to the source multilayer position to ensure the beam flux was at a maximum and uniform.

\begin{figure}[H]
   \begin{center}
   \begin{tabular}{c} %% tabular useful for creating an array of images 
   \includegraphics[height=8cm]{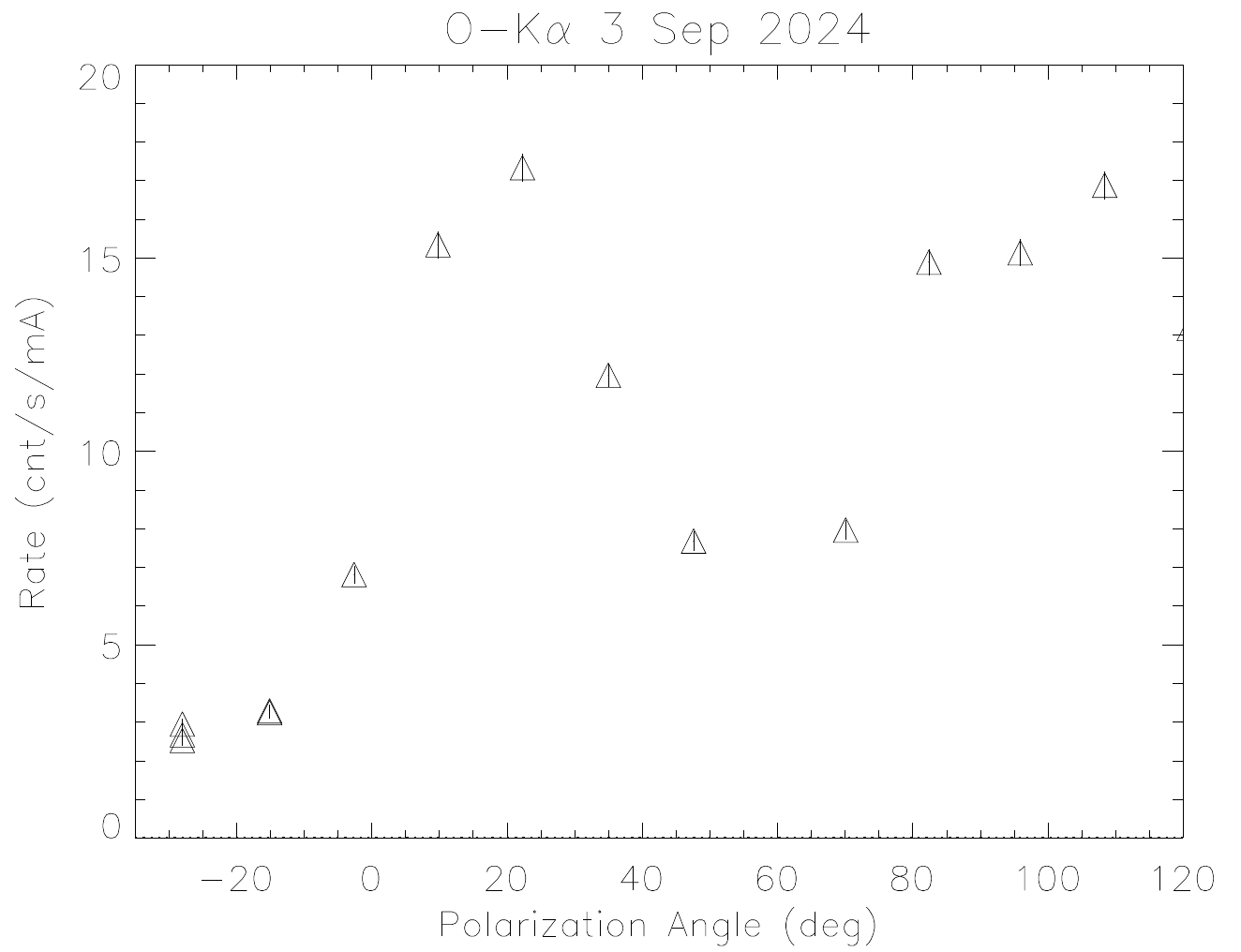}
   \end{tabular}
   \end{center}
   \vspace{-0.2in}
   \caption[example] 
%>>>> use \label inside caption to get Fig. number with \ref{}
   { \small \label{fig:oknormscan}
   Example of a set of measurements taken at different position angles of the polarized source.  An angle of zero was set by construction to the position where the source LGML is vertical, so the polarization of the beam is vertical.  While there are strong variations at this energy, they do not follow the pattern of a polarized source, which would have a 180$\degr$ period.
  }
   \end{figure}

We then obtained direct measurements of the beam at various source polarization angles by rotating the source about the beamline axis.  These data are important because they allow us to calibrate out any changes in the beam strength with polarimetry position angle, which arise from slight misalignments in the system. For carbon, we found that the beam was relatively constant with polarimetry angle, but all other anodes showed some variability with polarization angle.  A measurement of the direct beam count rate versus polarization for the O-K$\alpha$ line (sapphire anode) is shown in Figure \ref{fig:oknormscan}.  We normalized all multilayer modulation data with corresponding direct beam data.

Once these data were taken, we moved into the polarimetry mode configuration, shown in the lower panel of Figure \ref{fig:beamlineschematic}.  The camera must be warmed up and the beamline brought up to room pressure and opened to change configuration; the source multilayer is left in place so the uniformity calibration is unaffected.
%However, as the source and camera are turned off to make this change we are unable to take true reflectivity measurements of the multilayers.
In the polarimetry configuration, the beam illuminates most of the test multilayer.  It only reflects off of the locations on the multilayer where the energy of the input beam matches the Bragg peak energy of the multilayer (a vertical stripe).  This reflected stripe of light shines into the CCD, which is oriented vertically facing downwards towards the test multilayer.  We then perform a scan over several polarimetry angles, taking images of the reflected stripe of X-rays at each location.

\begin{figure}[H] 
   \begin{center}
   
   \includegraphics[width=15cm]{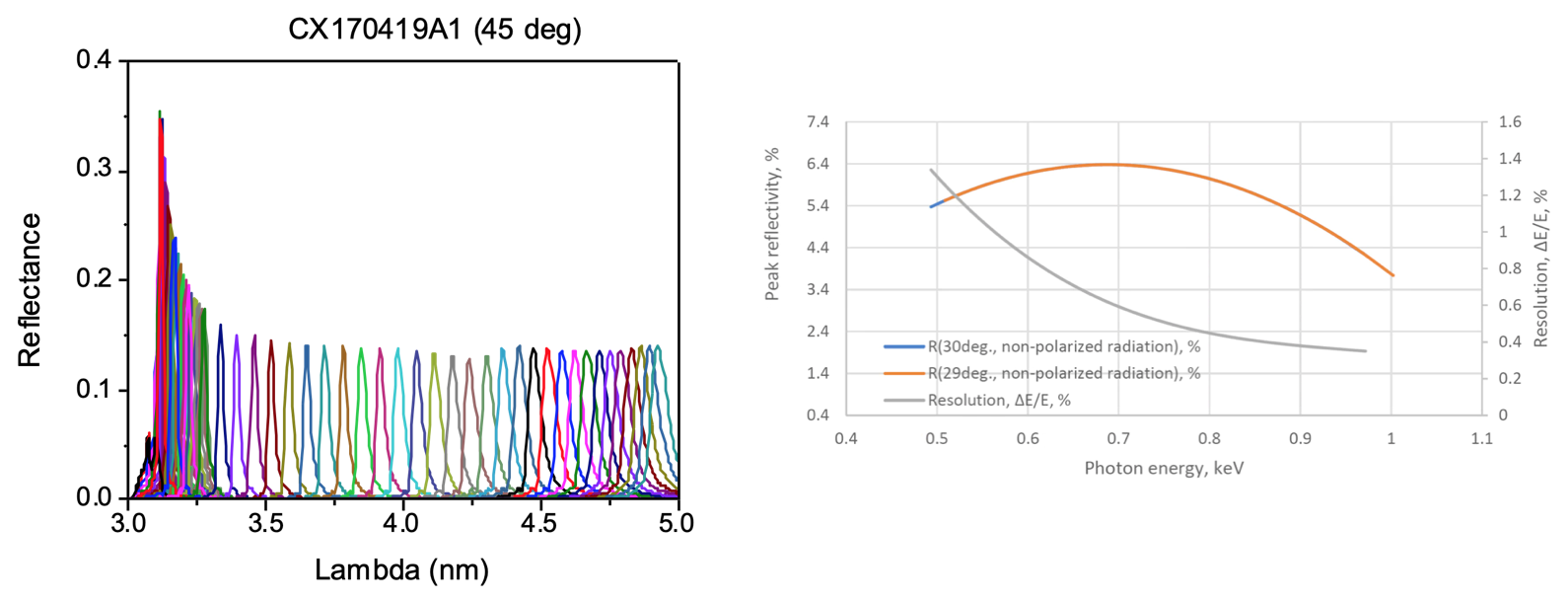}
   \end{center}
   { \label{fig:crsc-reflectivity} 
   \small  {\it Left:} Reflectance scans at regular locations on the Cr/Sc LGML, obtained at the Advanced Light Source. {\it Right:} Model of the reflectivity to unpolarized X-ray and spectral resolution of the W/B$_4$C LGML, provided by Rigaku.}
   \end{figure}

We tested two LGMLs in separate tests.  Each LGML was mounted on a rotation stage with the rotation axis perpendicular to the beamline at an orientation such that it aligned with the source LGML at a position angle of 90$\degr$ (as seen in Fig.~\ref{fig:beamlineschematic}).  One LGML consisted of 150 bilayers of Cr and Sc with a period gradient of 0.088 nm/mm over a range from 3.15 nm to 5.0 nm (about 250 to 400 eV).  At an incident angle of 45$\deg$, the reflectance (to predominantly polarized X-rays) was measured to to be 15 to 30\% at the Advanced Light Source (Fig. \ref{fig:crsc-reflectivity}).  It was tested at two energies: C-K$\alpha$ at 277 eV and Sc-L$\alpha$ at 395 eV.  At 45$\degr$, the Brewster angle, the polarization modulation was expected to be nearly 100\%.  The second LGML was purchased from Rigaku: a W/B$_4$C multilayer with 300 bi-layers designed to reflect light between 500 eV and 1 keV at a 30$\degr$ incidence angle with a linear period gradient of 0.027 nm/mm.  The reflectivity was modeled by Rigaku, peaking at 6.4\% for a 30$\degr$ incidence angle.
We tested the W/B$_4$C multilayer at 45$\degr$ at 395 eV and at 30$\degr$ at both O-K$\alpha$ at 525 eV and at Fe-L$\alpha$ at 705 eV.  For the 30$\degr$ measurements we changed the mounting location of the multilayer under test and rotated it perpendicular to the beamline to achieve the proper geometry.  The rotation angle was set using the system alignment laser, mounted near the X-ray source.

We expect as we rotate the polarization angle of the incoming light, to see a sinusoidal modulation of the strength of the light reflected off of the test multilayer.  For both multilayers measured at 45$\degr$, we expect there to be no reflection at all when the polarization angle is orthogonal to the surface of the test multilayer (100$\%$ modulation).  Because the incident light for the 30$\degr$ multilayer is coming in at a shallower angle than 45$\degr$ we expect its modulation factor to be lower, so we do not expect the reflected light signal to completely disappear.

\section{Results}
\label{sect:results}

\subsection{Cr/Sc Multilayer}
\label{sec:crsc}

\begin{figure}[H]
   \begin{center}
   \begin{tabular}{c} %% tabular useful for creating an array of images 
   \includegraphics[width=8cm]{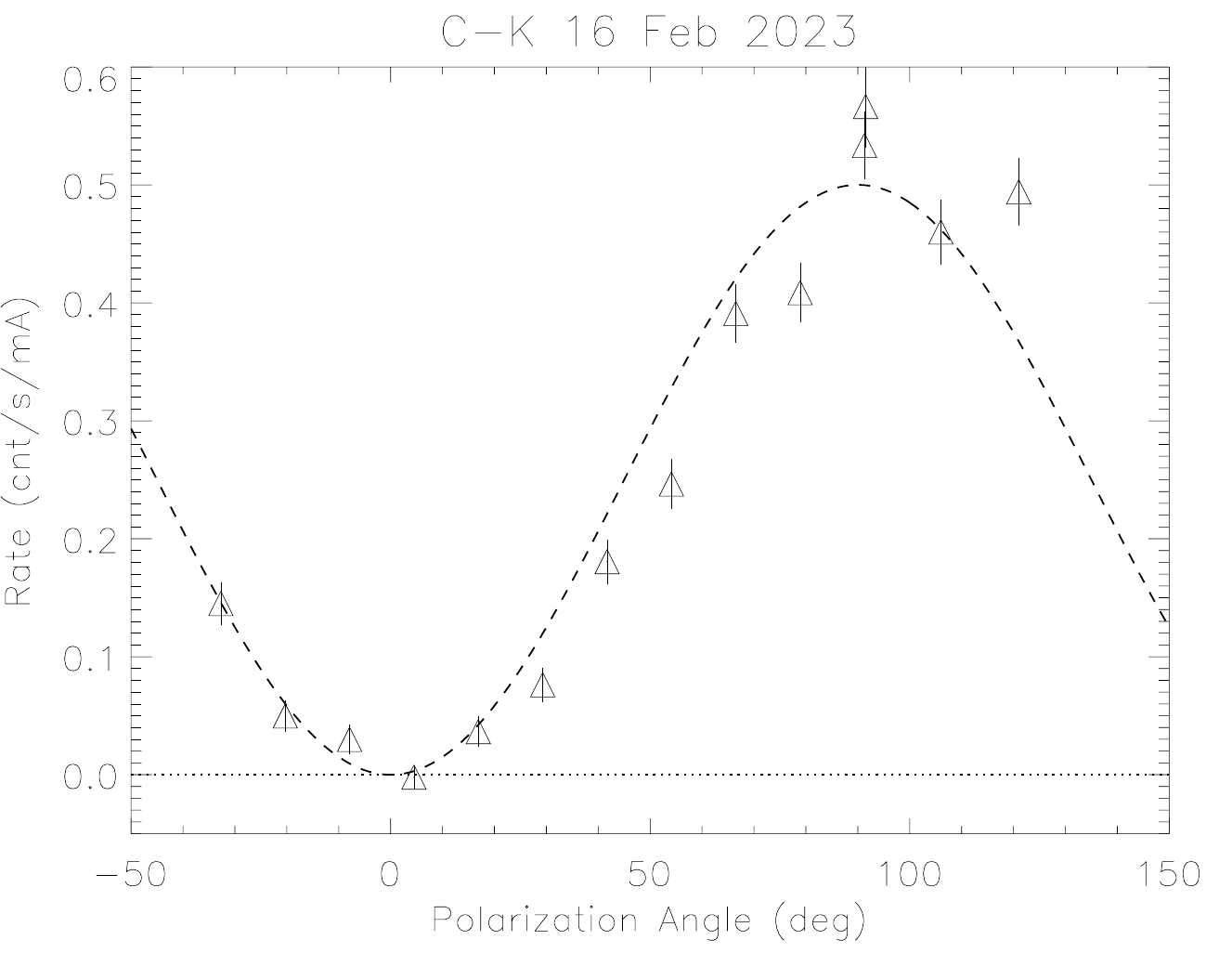} 
   \includegraphics[width=8cm]{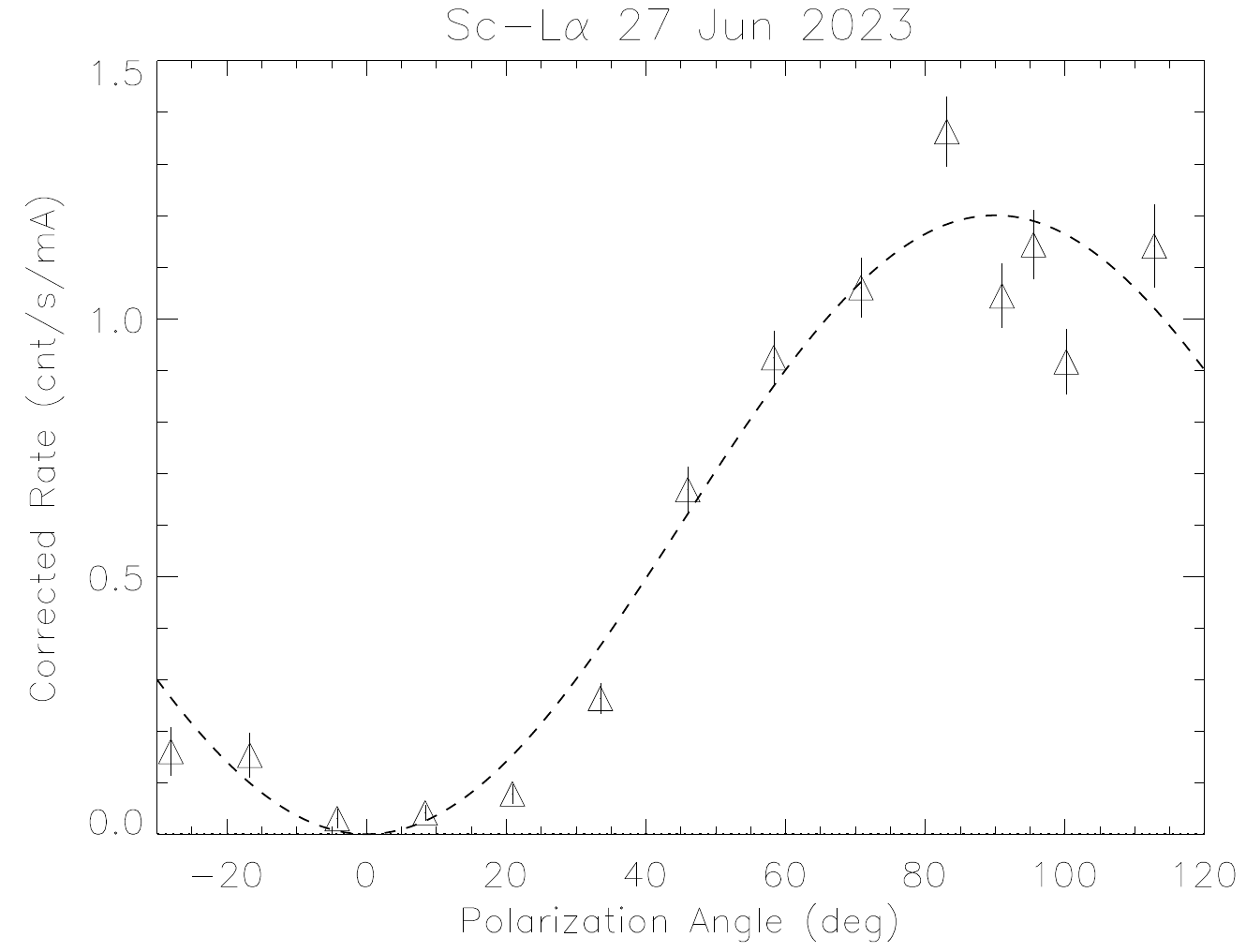}
   \end{tabular}
   \end{center}
   \vspace{-0.2in}
   \caption[example] 
%>>>> use \label inside caption to get Fig. number with \ref{}
   { \small \label{fig:crsc_results}
   Polarimetry results for the Cr/Sc LGML, measured at two energies.  The modulation is indistinguishable from 100\% in both cases.
  }
   \end{figure}

As mentioned earlier, we tested the Cr/Sc multilayer at 45$\degr$ incidence at both C-K$\alpha$ (277 eV) and Sc-L$\alpha$ (395 eV).  These two tests were made without breaking configuration.  To change between the energies we changed the anode on the source carousel and moved the source multilayer.  The X-rays reflecting from the test multilayer appeared in a different location on the detector because light was reflecting off the multilayer location corresponding to the Bragg condition for the selected energy.  The results are shown in Figure \ref{fig:crsc_results}.  We see 100\% modulation, as the signal goes completely to zero in both cases at 0$\degr$ source rotation, as designed.  This functions as a proof of concept that the laterally graded multilayer mirror paired with an X-ray detector functions as a broad-band soft X-ray polarization detector and also demonstrates that the source produces 100\% polarized X-rays.

\subsection{W/B$_4$C Multilayer}

\begin{figure}[H]
   \begin{center}
   \begin{tabular}{c} %% tabular useful for creating an array of images 
   \includegraphics[width=8cm]{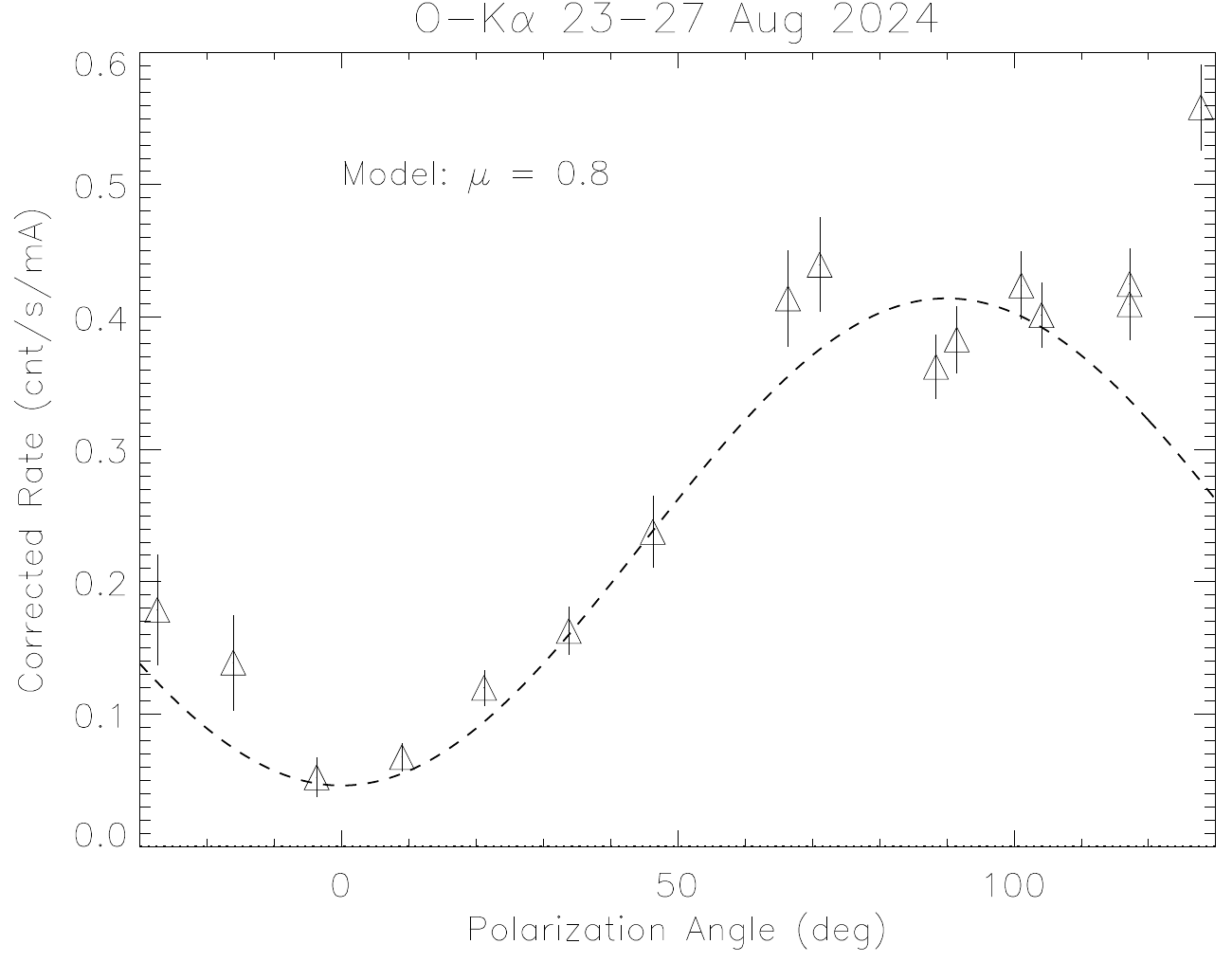} 
   \includegraphics[width=8cm]{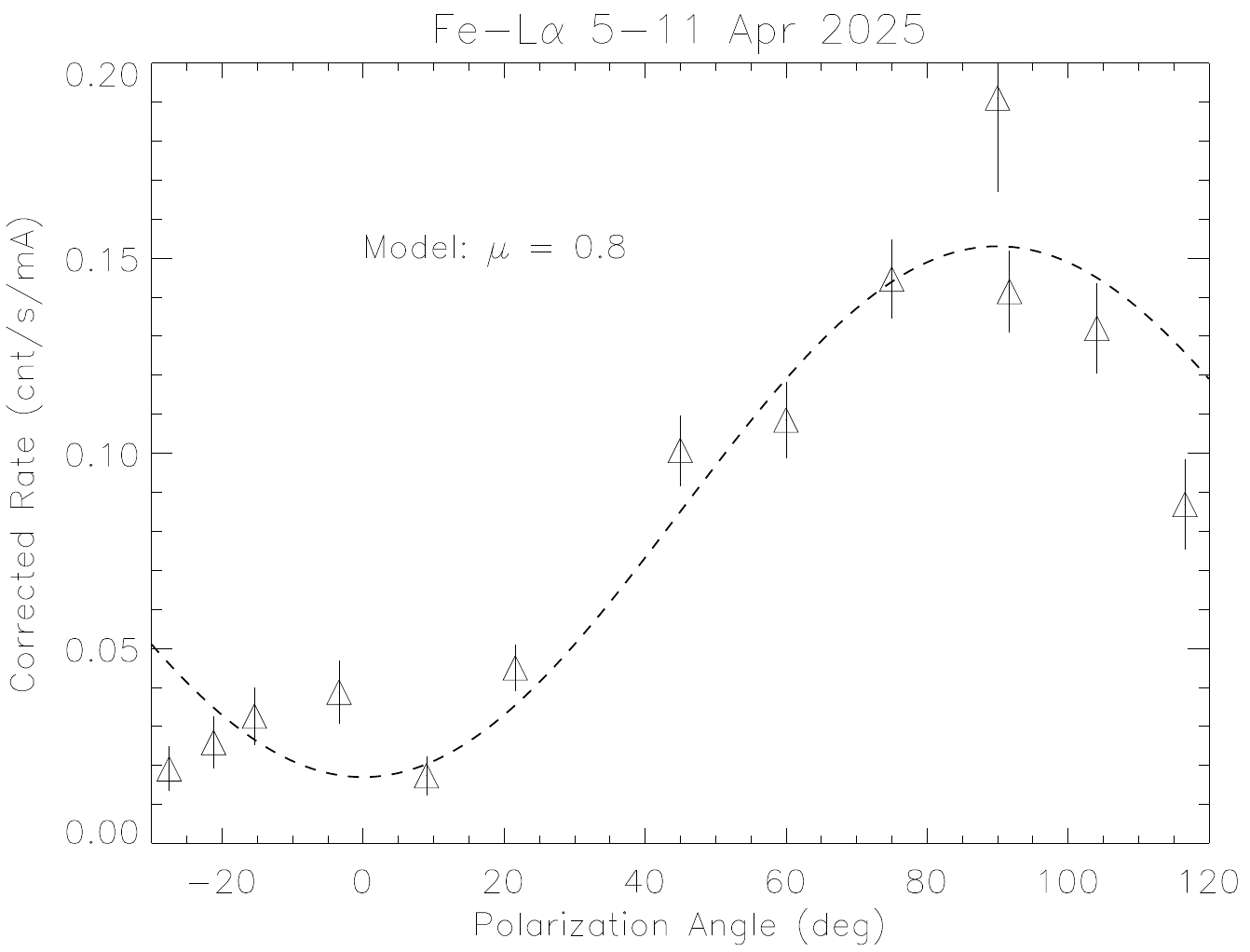}
   \end{tabular}
   \end{center}
   \vspace{-0.2in}
   \caption[example] 
%>>>> use \label inside caption to get Fig. number with \ref{}
   { \small \label{fig:wb4_30_results}
   Polarimetry results for the W/B$_4$C LGML, measured at 30$\degr$ incidence and two energies.  The modulation is about 80\% in both cases.
  }
   \end{figure}

For the W/B$_4$C multilayer, we measured using O-K$\alpha$ (525 eV) and Fe-L$\alpha$ (700 eV) to test its function over part of its designed 0.5-1 keV band.
%In its current configuration, 700 eV is the highest monochromatic energy we are able to produce in the beamline.
We do not expect 100\% modulation as we are operating short of Brewster's angle.  The results of the scans at these two energies are shown in Figure~\ref{fig:wb4_30_results}.
We find that the modulation is roughly 80\% in both cases.

\begin{figure}[H]
   \begin{center}
   \begin{tabular}{c} %% tabular useful for creating an array of images 
   \includegraphics[width=8cm]{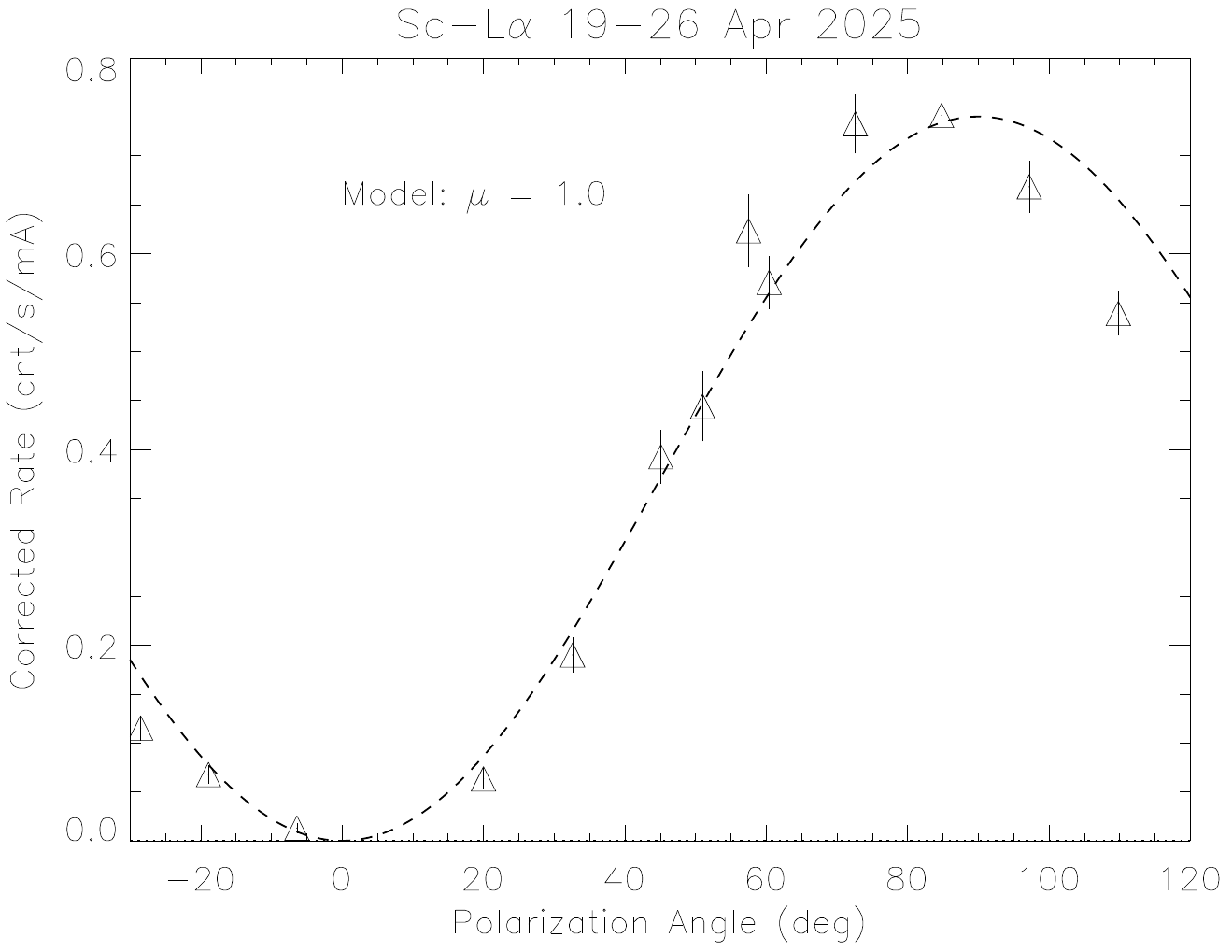}
   \end{tabular}
   \end{center}
   \vspace{-0.2in}
   \caption[example] 
%>>>> use \label inside caption to get Fig. number with \ref{}
   { \small \label{fig:wb4_45_results}
   Polarimetry results for the W/B$_4$C LGML, measured at 45$\degr$ incidence at 395 eV.  The modulation is about 100\%, as expected.
  }
   \end{figure}

As the incidence angle increases, the Bragg peaks move to lower energies.  To test how the WB$_4$C multilayer performed at Brewster's angle, we tested it at 45$\degr$ with the Sc-L$\alpha$ line (395 eV).  The results of this scan are shown in figure \ref{fig:wb4_45_results}.  In this case, the modulation is indistinguishable from 100\% modulation because the mirror is used at Brewster's angle.  

\section{Conclusions and Future Work}
With these measurements, we have provided proof of concept that laterally graded multilayer mirrors can be used for spectropolarimetry measurements over a broad band.  The Cr/Sc mirrors being used for REDSoX can be used at Brewster's angle to provide 100\% polarimetry modulation in the 0.2-0.4 keV band.  We also showed that the W/B$_4$C mirrors can be used at 30$\degr$ to make such measurements in the 0.5-1 keV range.  Future work will include investigations of other material combinations that might provide higher reflectivities in the higher energy band.

% \disclosures 
\subsection*{Disclosures}
The authors declare no relevant financial interests or potential conflicts of interest related to this manuscript.

\subsection* {Acknowledgments}
This work was suppported by NASA APRA grant number 80NSSC20K1012.  The authors also acknowledge support from the MIT Kavli Institute's Research Investment Fund. 

%%%%% References %%%%%

\bibliography{main,polarimetry24}   % bibliography data in report.bib
\bibliographystyle{spiejour}   % makes bibtex use spiejour.bst

\end{document}